# Cultural Cartography with Word Embeddings

Dustin S. Stoltz
*Lehigh University*

Marshall A. Taylor
*New Mexico State University*

Draft on 4/6/2021

## Abstract[1]

Using the frequency of keywords is a classic approach in the formal analysis of text, but has the drawback of glossing over the relationality of word meanings. Word embedding models overcome this problem by constructing a standardized and continuous "meaning space" where words are assigned a location based on relations of similarity to other words based on how they are used in natural language samples. We show how word embeddings are commensurate with prevailing theories of meaning in sociology and can be put to the task of interpretation via two kinds of navigation. First, one can hold terms constant and measure how the embedding space moves around them—much like astronomers measured the changing of celestial bodies with the seasons. Second, one can also hold the embedding space constant and see how documents or authors move relative to it—just as ships use the stars on a given night to determine their location. Using the empirical case of immigration discourse in the United States, we demonstrate the merits of these two broad strategies for advancing important topics in cultural theory, including social marking, media fields, echo chambers, and cultural diffusion and change more broadly.

[1] We would like to thank Lisa Kressin, Omar Lizardo, John Levi Martin, Terence E. McDonnell, Laura Nelson, David G. Ortiz, Michael Lee Wood, Andrea Voyer, the participants of the Workshop on Computational Text Analysis in the Social Sciences at Linköping University, Norrköping, Sweden, and the participants of the Workshop on Big Data Applications, Challenges, and Techniques at New Mexico State University, Las Cruces, NM, USA, for comments on earlier drafts and work contributing to this paper. We are grateful for the feedback from the editors, Craig Rawlings and Clayton Childress, as well as two anonymous reviewers. We would also like to thank Andrea Voyer for suggesting star metaphors to explain word embeddings, as well as all of the Twitter users who pointed us toward many word embedding papers we now reference. A replication repository for this paper is available at: https://github.com/dustinstoltz/cartography_poetics.

Meaning is often central to cultural analysis (Mohr et al., 2020, p. 2; Spillman, 2020, p. 1). As texts offer widely-available and unobtrusive sources of "meaning in the wild," formal text analysis has steadily produced a suite of tools for studying how meaning is articulated by individuals, groups, and organizations (Mohr, 1998; Mohr & Duquenne, 1997; Mohr & Lee, 2000). Many of these tools, however, are based on token counting, which is generally a poor fit with the relational theories of meaning they accompany.

The meaning of texts (e.g., books, articles, or social media comments) is, typically, established by the extent it references certain concepts or entities (Weber, 1984), often by *counting* certain words (Namenwirth & Weber, 2016; Weber, 1990). Lee and Martin (2014) refer to this process as *cultural cartography* in that, like a topographic map of terrain, it selectively simplifies texts in useful ways. The main problem with counting approaches is not that word order or subtlety is lost, or that certain words are selected as representative (cf. Breiger et al., 2018; Mohr et al., 2015). Indeed, "it is precisely because of their impoverishment that maps are useful" (Lee & Martin, 2014, p. 12). *Rather, procedures based on counting tokens are appropriate for discrete measures (i.e., either-or) but not graded measures (i.e., more or less).* The spread of populist ideologies across political campaigns, the prevalence of diversity rhetoric among management consultants, or the relative difference between labor discussions in Germany, Iceland, and the United Kingdom are all cases where graded measures are more appropriate than discrete measures for analyzing meaning. Each case refers to generic ideas that can be more and less present or discourses that can be more and less alike. Concerns about the *magnitude* of conceptual engagement and similarity are central to cultural analysis; but count-based measures are ill-suited for measuring magnitudes.

The alternative we propose remains within the spirit of Lee and Martin's cultural cartography in that we aim to simplify texts in faithful ways while preserving the graded, relational meanings of words (Kirchner & Mohr, 2010; Mohr, 1998). Word embedding models offer a means to do just this. These models allow us to substitute the comparison of frequencies with the comparison of distances by providing standardized maps of meaning space in which semantic relations are realized as spatial relations. Importantly, these models do not "understand" meaning, nor do these procedures substitute for interpretation (Chakrabarti & Frye, 2017; Nelson, 2020; Popping, 2012).[2] Instead, they "condense information to facilitate an intersubjectively valid interpretation" (Lee & Martin, 2014, p. 1).

In what follows, we overview word embedding models as a means of mapping meaning space. We then organize the various applications of embeddings in cultural analysis into two broad kinds of "navigation." The first we refer to as *variable embedding space* methods, which involve holding terms constant to measure how the meaning space moves around them. Much like astronomers measured the changing relative locations of celestial bodies with the seasons, these methods measure the changing relative locations of words. More technically, these methods entail splitting a corpus by a covariate—e.g., time periods or authors—and training word embeddings on each subset of the corpus. Second, *fixed embedding space* methods use the same map of meaning space to measure how documents[3] or authors[4] change in relation to it. Here, just as ships determine their location relative to stars in the night sky, these methods measure the relative position of authors or documents *vis-à-vis* a single set of word embeddings. Documents or authors are represented as nebulae or clouds of locations, and measuring similarities to other documents, authors, words, or concepts becomes a transportation problem. We illustrate both approaches using the case of immigration discourse and its evolution in the United States.

---

[2] It is a category error to say word embedding models "understand" meaning (Bender & Koller, 2020; Glenberg & Mehta, 2009), and an example of the "symbol grounding" problem (Harnad, 1990; Lizardo, 2016).

[3] We use "document" to refer to any aggregate of text: books, articles, chapters, paragraphs, social media comments, etc.

[4] We use "author" to refer to individuals, groups, communities, or organizations responsible for producing texts.

# Mapping Meaning Space

## LANGUAGE MODELING AND RELATIONAL THEORIES OF MEANING

There are two strategies for representing words as numbers: discrete and distributed. In the simplest discrete approach, each unique word is represented by a binary vector where each unique word in the corpus is assigned to a position in the vector (i.e., one-hot encoding). Comparing whether two terms (tokens) are the same entails measuring whether their term (type) vectors have a "1" in the same position. Words are either precisely the same, or they are not.[5] The relationship between the numbers standing in for terms, however, do not correspond to the graded relationship between the meanings of terms (Smith, 2019, p. 2). *Two terms that mean similar things will be just as different as two terms that mean dissimilar things.* The second strategy overcomes this problem by creating "distributed" representations of terms' meanings. Word embedding models induce such "distributed" representations by summarizing the co-occurrences of words in natural language corpora (Lenci, 2018).

---

[5] Each unique word (type) could be assigned an integer at random, where word (tokens) are the same if their assigned integers are precisely equal. The problem remains: words are either exactly the same, or not.

Chomsky famously asserted that corpus linguistics and the statistical study of text was a dead-end (Chomsky, 2002, pp. 15–20; R. Harris, 1995, pp. 96–98),[6] and yet the kernel of the "distributional hypothesis" (Sahlgren, 2008) is found in the work of Chomsky's dissertation advisor, Harris. Drawing on Bloomfield and Sapir in particular, Harris argued (1954, p. 156) "difference of meaning correlates with difference of distribution" (see also Joos, 1950).[7] Similarly, Firth—influenced by Malinowski (Firth, 1935; Rose, 1980; Young, 2011)[8]—stated that words' meanings can be inferred from their "habitual collocations":

> …a text in such established usage may contain sentences such as 'Don't be such an ass!', 'You silly ass!', 'What an ass he is!' In these examples, the word ass is in familiar and habitual company, commonly collocated with *you silly-*, *he is a silly-*, *don't be such an-*. You shall know a word by the company it keeps! One of the meanings of ass is its habitual collocation with such other words as those above quoted. (Firth, 1957, p. 11, emphasis in original)

This basic hypothesis was supported by early statistical analyses (Henley, 1969; Rubenstein & Goodenough, 1965), leading Miller and Charles (1991, p. 24) to foreshadow word embeddings: "the general idea is to consolidate various kinds of information about a word's contexts into a single representation that characterises those contexts." The goal, then, is to assign each word a *single vector* such that the "gradedness in distributional representations correlates with gradedness in semantic phenomena" (Boleda, 2020, p. 228).[9]

---

Inferring a word's meaning by summarizing its (linguistic) context aligns with relational theories of meaning—pioneered by John Mohr (Kirchner & Mohr, 2010; Mohr, 1998, 2000), along with Pierre Bourdieu, Ron Breiger, Ann Mische, Harrison White, Viviana Zelizer and many others (Mische, 2011; Pachucki & Breiger, 2010; Zelizer, 2012)—as well as pragmatic, embodied, connectionist, and practice-theoretic approaches to cultural learning (Arseniev-Koehler & Foster, 2020; Ellis, 2019; Erk, 2016; Glenberg & Robertson, 2000; Ignatow, 2007, 2016; Landauer & Dumais, 1997; Lizardo et al., 2019; Strauss & Quinn, 1997).

Accordingly, social scientists are beginning to use word embeddings (e.g., Boutyline et al., 2020; Hofstra et al., 2020; Jones et al., 2020; Linzhuo et al., 2020; Ornaghi et al., 2019; van Loon & Freese, 2019). This early work largely builds on the fact that word embeddings mirror the stereotypical racial, ethnic, and gender related biases found in the texts they are trained on (Brunet et al., 2018; Caliskan et al., 2017; Lewis & Lupyan, 2020).[10] While this can be a concern for some downstream applications—indeed, leading some researchers to attempt to "debias" word embeddings (Bolukbasi et al., 2016a; Gonen & Goldberg, 2019)—this is a strength for those studying these associations as features of the social world. Rather than distortions in the semantic

---

1954, p. 156 emphasis added). This is contrary to problematic neo-Saussurean formulations in sociology (Stoltz, 2019), wherein words' meanings are said to be *entirely constituted by* their linguistic context alone. See Bender and Koller (2020, p. 7) for a similar critique of Wittgensteinian formulations: "the slogan 'meaning is use'... refers not to 'use' as 'distribution in a text corpus' but rather that language is used in the real world to convey communicative intents to real people."

[10] Although an open discussion, these models are found to accurately reflect co-occurrence patterns in corpora (Ethayarajh et al., 2019, p. 9), as opposed to "exaggerating" associations. In particular, researchers have validated these associations against traditional techniques, such as surveys and implicit association tests (Joseph & Morgan, 2020; Kozlowski et al., 2019).

space, these associations reflect the contours of cultural formations. Some key terms relating to word embeddings and their definitions are presented in Table 1.

**[TABLE 1. HERE]**

# WORD EMBEDDINGS: SOME TECHNICAL DETAILS

## The Term-Co-occurrence Matrix

The simplest distributional model begins by assigning each unique term in a corpus a vector. Each unique term is also a "position" in that vector indicating co-occurrences. Each entry in a term's vector is the frequency it occurs "next to" the term corresponding to that dimension within a given window, say five terms on either side.[11] The result is a target-term by context-term matrix, and the cells indicate the frequency a term (row) co-occurs with another term (column) within that window (see also Lee & Martin, 2014, pp. 15–16). This matrix is often called a *term-co-occurrence* or *term-context matrix* (TCM), and formalizes Firth's "habitual collocations" (1957, p. 11).

## Dimension Reduction

There are two drawbacks to using the TCM alone. First, using larger corpora, each term's vector would be long (as long as the number of unique terms in the corpus) and sparse (with many zeros indicating two words never co-occur). Second, and more importantly, two target terms are considered similar to the extent they share an exact matching set of co-occurring terms.

---

[11] Goldberg (2016, pp. 367–369) provides a more detailed discussion: "the size of the sliding window has a strong effect on the resulting vector similarities. Larger windows tend to produce more topical similarities (i.e."dog", "bark" and "leash" will be grouped together, as well as "walked", "run" and "walking"), while smaller windows tend to produce more functional and syntactic similarities (i.e."Poodle", "Pitbull", "Rottweiler", or "walking", "running", "approaching")."

For both these reasons, dimension reduction is applied to the TCM (Lenci, 2018, p. 157; Wong et al., 1985). The TCM is reduced not by selecting "columns" (i.e., feature selection), but rather by finding a few latent patterns summarizing the information in the matrix (i.e., feature extraction). This is motivated by the idea that "co-occurrences collected from corpora are noisy data that hide more abstract semantic structures" (Lenci, 2018, p. 157).

To summarize, at their most basic, word embedding models involve creating a TCM and reducing its dimensions. The *n*-dimensional TCM is "reduced" to at least *n*-1 latent dimensions and thus those original elements have been *embedded* into a "lower-dimensional" space. Advances in embedding models primarily improve the "tuning" of this low-dimensional mapping (see Appendix B). This notion forms the basis of "embedding" methods in computer science and is a mathematical feature common to many matrix factorization tools familiar to social scientists—e.g., factor analysis and correspondence analysis.[12] In comparison to other kinds of dimensional analysis, word embeddings have a relatively high number of dimensions because, in part, relational meanings of words are *intransitive* (word *A* can be similar to word *B*, and word *C* can be similar to word *B*, but this does not necessarily entail word *C* being similar to *A*). This necessitates higher dimensionality to prevent distortions.

## From Semantic Relations to Spatial Relations

Next, and key for our argument, this matrix can be interpreted *geometrically*, such that the row vectors designate terms' locations in a continuous space (usually

---

[12] Methods falling under the labels "topic modeling" and "latent semantic analysis" involve applying dimension reduction on a DTM (as opposed to the TCM). The dimensions of embedding matrices are not "interpreted" directly (cf. Bodell et al., 2019), as one might interpret "topics" or "principal components," but rather the relations between elements' locations as defined by those dimensions.

Euclidean). As geometric relations correspond to semantic relations, we can use operations from linear algebra to extract meaningful associations (Erk, 2012). Mainly, distance measures *semantic similarity* between two words, but also adding, subtracting, and averaging vectors produce intuitive results—sometimes referred to as relation induction, relation extraction, or semantic projection—useful for the cultural analyst (e.g., Arseniev-Koehler & Foster, 2020; Kozlowski et al., 2019; Nelson, 2021; Taylor & Stoltz, 2020). For example, subtracting "man" from "king" and adding "woman" produces a vector near "queen," or, adding "south" to "Africa" will be near the intersection of the two terms' co-occurrence contexts and thus surrounded by words related to "South Africa." While there are other kinds of relations one could extract  (see the Discussion), we focus on similarity.

## Pretrained Embeddings and the Selection of Training Corpora

[TABLE 2. HERE]

Selecting corpora with which word embeddings (hereafter just *embeddings*) will be trained is an important consideration. Pretrained embeddings (see Table 2) are estimated using large-scale, widely-representative, "naturally occurring" corpora. For example, fastText embeddings are trained on several billion web pages from the Common Crawl (Joulin et al., 2016). More targeted pretrained embeddings also exist, such as those trained on the Corpus of Historical American English separately for each decade from 1810 to 2000 (Davies, 2012).

We contend (along with Spirling & Rodriguez, 2019) that researchers should *default* to using pretrained sources, and use corpus-trained if required by the research question.[13] The computational resources and time required to accurately train these models can be borne once (Lazer & Radford, 2017, p. 33), and pretrained embeddings enhance comparability across studies. There are, of course, limitations to pretrained embeddings (cf. Bender et al., 2021). In particular, they may not be available for a specific context or language community, or use training corpora that systematically omit relevant populations.

Corpus-trained (or locally-trained) embeddings, by contrast, are word vectors trained exclusively on the researcher's corpus, or a broader corpus comprising the researcher's specific corpus (e.g., exploring the works of Marx using embeddings trained on all 19th century social science books). In general, a researcher will use pretrained embeddings if interested in analyzing how their documents relate to widely shared semantic associations within a given language community (i.e., "mainstream" culture). A researcher may want to train their own embeddings if they instead want to analyze any semantic idiosyncrasies within their targeted collection (e.g., Nelson, 2021), are attempting to tap into "counter-cultural" meanings, or if pretrained embeddings are not available for a specific language community, context, or population.

## Navigating Meaning Space

Interpretation proceeds by measuring how units of analysis are related to each other by analogy to space. We can interpret, say, what it means for "immigration" to be a certain distance from "school" when we compare this to its distance from "family." That is, a *fixed waypoint* must be defined from which we gain our perspective on the relative

---

distances of other points. Accordingly, we divide methods based on what unit of analysis is "fixed" (see Table 3).

In our first group of methods, the terms are fixed while the relations between them are allowed to vary. We refer to this as *variable embedding space methods*. We accomplish this by subsetting our corpus by a covariate, usually time or author, training multiple sets of embeddings and measuring the differences in the relative location of key terms within each of these spaces—much like early astronomers measured how the positions of celestial bodies changed across the seasons. In our second group—which we call *fixed embedding space methods*—the embedding space is held constant while measuring how documents or authors differ in relation to each other. We accomplish this by measuring the relative locations of documents or authors defined as aggregates of terms in a single set of embeddings—just as ships use the stars at a given time to determine their location.

**[TABLE 3. HERE]**

## VARIABLE EMBEDDING SPACE

Kulkarni et al. (2015) offered one of the first studies to fix terms and compare their changing relative position in embeddings trained on text from different time periods. They empirically demonstrate the well-known discursive shift over the 20th century where the word "gay" changed from being located beside "cheerful" and "frolicsome" to being near "lesbian" and "bisexual." Similarly, Garg et al. (2018) track changes in gender and ethnic biases in English over the same time period by comparing the changing distances between gender- and ethnicity-related terms and a list of adjectives and occupational terms (see also Jones et al., 2020; Kozlowski et al., 2019). This involves a "time-lapse" approach, training on texts from different time periods and measuring changes in the meaning space between periods.

We could subset a corpus by variables other than time, and train separate models on each subset to measure how the meaning of terms varies across, e.g., individuals, communities, or organizations. For example, Bonikowski et al. (2019) divide presidential candidates' speeches by individual candidates to find the 50 nearest neighbors of two focal terms derived from each candidates' embeddings. As each model is trained on separate candidates, the associations are specific to the candidate. Similarly, Zannettou et al. (2018) compare models trained on text from different communities: the 4-chan board /pol/ and Gab (see also An et al., 2019; Rho et al., 2018; Schild et al., 2020).

There is a caveat to comparing embeddings trained on separate corpora. Matrices may need to be "aligned" (Hamilton et al., 2016, p. 4; Kulkarni et al., 2015, p. 4) because the random processes involved render non-unique solutions even when training on the same corpora with the same methods (cf. Orlikowski et al., 2018; Tahmasebi et al., 2018). Simply put, alignment involves rotating and scaling of two or more matrices, preserving the distances between terms within each embedding while approximately aligning the terms across embeddings (Artetxe et al., 2016; Ruder et al., 2019).

## Measuring The Distance Between Terms Over Time

To demonstrate variable embedding methods, we first show the simplest use, which is measuring how the distance between two (or more) words differs over time. Specifically, we measure how the meaning of "immigration" has shifted in American

English using embeddings trained on the Corpus of Historical American English for each decade from 1880[14] to 2000 (Davies, 2012; Hamilton et al., 2016).[15]

Here, we take the cosine similarity between the vector of "immigration," on one hand, and the vectors for "job," "crime," "family," and "school," on the other, for each decade. In the plot (see Figure 1), immigration was strongly associated with "job" at the end of the 19[th] century, and while "school," "family," and "job" slowly increased in the 20[th] century, immigration has grown even more associated with "crime." While "immigration" had its lowest observed cosine similarity with "crime" in the 1890s (0.076, on a scale of -1 to 1, where larger positive values indicate more similarities in discursive context), it reached its peak similarity in the 1990s (0.344), with the nearest other similarity with "immigration" in that decade being school (0.087). These temporal trends reflect a change in the meaning space surrounding immigration in American English over nearly one and half centuries. Culture researchers might then assess the extent to which variance in these discursive shifts for a population over time are explained by period- or cohort-level effects—a linguistic extension of the cultural change research carried out mostly using survey data (cf. Hamilton et al., 2016; Kiley & Vaisey, 2020; Vaisey & Lizardo, 2016).

**[FIGURE 1. HERE]**

## Measuring The Distance Between Terms and Cultural Dimensions

Next, we extract a semantic direction pointing toward a pole of the cultural dimensions of race, social class, and morality. This involves identifying sets of juxtaposing terms for each cultural dimension—understood as generic binary oppositions that "individuals use in everyday life to classify agents and objects in the world" (Kozlowski et al., 2019, p. 911). For example, for social class, this would be affluence vs poverty, rich vs poor, and so on. The vector for "poverty" is subtracted from "affluence." This is repeated through the set of class-related pairs, averaging the results. The final vector represents one pole of the cultural dimension of affluence.[16] The affluence, race, and morality dimensions were constructed using term pairs taken from Kozlowski et al. (2019, pp. 935–937).[17]

---

[14] The first valid word vector for "immigration" is in 1880 and thus this is our starting point.

[15] https://nlp.stanford.edu/projects/histwords/

We measure the changing position of the term "immigrant" as well as "citizen" relative to these cultural dimensions in each decade. The *y*-axis of Figure 2 (both panels) shows that, regardless of decade, "citizens" (yellow triangles) tend to be closer to the "white" pole of the race dimension, whereas "immigrants" (purple dots) tend to be closer to the "black" pole. In the *x*-axis of the left panel, we see that "citizens" in 2000 is closer to the "high class" pole of the affluence dimension than is "immigrants" in 2000. In the *x*-axis of the right panel, we see that "citizen" has clustered near the "good" pole of the morality dimension throughout the decades, whereas "immigrants" have generally moved closer to "good" over the decades, with 1980, 1990, and 2000 being the closest to "good."

These associations could be interpreted in one of two (non-mutually exclusive) ways. First, engagement with the "citizen," "white," "high class," and "good" poles of their respective cultural dimensions may be correlated because—unlike "immigrant," "black," "low class," and "bad"—these concepts have been both largely *absent* in U.S. public discourse over time. This interpretation suggests that these "discursively absent" poles have been remarkably stable *unmarked* (and therefore taken-for-granted and normative) categories for citizenship, race, social class, and morality discourses in the United States (Brekhus, 1998).[18] The second interpretation is that the temporal stability in how the immigrant-citizen dimension correlates with these other dimensions reflects durable symbolic boundaries in the U.S. This corroborates prior research finding the "immigrant-as-nonwhite"[19] vs. "citizen-as-white" has been an enduring cultural structure in U.S. public discourse and attitudes (Mora & Paschel, 2020; Sáenz & Douglas, 2015).

---

[16] There are several procedures for deriving a "semantic direction" from an embedding space (Arseniev-Koehler & Foster, 2020, pp. 18–9; Bolukbasi et al., 2016b, pp. 42–3; Boutyline et al., 2020; Ethayarajh et al., 2019; Kozlowski et al., 2019, p. 943 fn 8; Larsen et al., 2015, p. 5; Taylor & Stoltz, 2020).

[17] The embeddings did not have the following for all time periods: advantaged, propertied, sumptuous, swanky, ritzy, uncorrupt, pureness, necessitous; skint, penurious, unmonied, unprosperous, moneyless, transgressive, knavish, afro.

**[FIGURE 2. HERE]**

## FIXED EMBEDDING SPACE

A straightforward use of a fixed embedding space is measuring the relationship between key terms or cultural dimensions. Arseniev-Koehler and Foster (2020), for example, train a model on over a hundred thousand *New York Times* articles and measure the distance between terms related to "obesity" and key cultural dimensions: gender, morality, health, and socioeconomic status (see also Nelson, 2021). An analyst can also use fixed embeddings to compare the relations between documents or authors, in particular determining the general *semantic similarity* between documents or authors.

---

[18] A more in-depth analysis with the "marked-unmarked" framework might benefit more from a keyword absence/presence method, since a category's "unmarked" status is, by definition, signified by its absence.

[19] A more systematic analysis would either construct a "nonwhite-white" semantic direction (as opposed to "black-white"), or construct several semantic directions with "white" as one pole and as series of non-white racial/ethnic categories as the opposite poles. These steps would be necessary to more confidently link this finding to the "immigrant-as-nonwhite" symbolic boundary interpretation.

These similarities can be used for a variety of ends such as comparing document revisions, text classification, measuring content change, building semantic networks, or studying content diffusion (e.g., Ahlgren & Colliander, 2009; Berry & Taylor, 2017; Strang & Dokshin, 2019; Teplitskiy, 2016; Zhang & Pan, 2019).

## Document-by-Document Distances

Prior social scientific research estimating document similarities often relied on discrete word representations. For example, Bail (2012) used plagiarism detection software (Bloomfield, 2008) to compare press releases by civil society organizations to media coverage. Similarly, Grimmer (2010) used the same software to compare press releases from Senate offices to media coverage. This software, after lemmatizing two documents, searches for exact matches in strings of six words. This technique, then, treats words as either the same or not, therefore not accounting for the graded, relational meaning of words.

Farrell's (2016) method is even closer in spirit to the fixed embeddings approach. To compare the similarity of writings by climate contrarian organizations with news outlets and political offices, Farrell reduced the dimensionality of the DTM by applying SVD, and comparing the cosine similarity of the resulting document vectors (Deerwester et al., 1990). While this formed the backbone of information retrieval for decades, recent research demonstrates that similarity measures using embeddings, specifically Word

Mover's Distance (Kusner et al., 2015),[20] outperform this procedure on various tests, including plagiarism detection (Tashu & Horváth, 2018).

With Word Mover's Distance (WMD), documents are represented not just by counts of unique terms, but also the semantic relations between those terms. A document becomes a cloud of locations in the embedding space. Determining similarity is treated as a transportation problem where the "cost" of moving one document's cloud of locations to another is equivalent to the semantic similarity between two documents. The DTM provides the "amounts" to be moved and the embedding matrix provides the "distances" that these amounts are moved. The result is a document-by-document similarity matrix, or an author-by-author similarity matrix if similarities are averaged or each document has a single author: for example, Pomeroy et al. (2019) use this to measure the relationship between nation-states based on the similarity of their United Nations speeches.

To demonstrate WMD (using LC-RWMD[21]), we find the document-by-document similarity matrix between a preprocessed (see Appendix A) corpus of U.S. news articles[22] and a preprocessed corpus of press releases published by immigration-focus advocacy organizations.[23] We collected a total of 986 press releases from two far-right organizations, the Center for Immigration Studies ($N = 160$) and the Federation for American Immigration Reform ($N = 379$), as well as two left-wing organizations, National Network for Immigrant and Refugee Rights ($N = 119$) and Women's Refugee Commission ($N = 328$). The publication date of the press releases ranged from 1998 to 2020 (median of 2015), but they were "pooled" together and compared to news articles across time points ranging from January 2016 to July 2017. We select three news organizations considered more right-leaning (*Breitbart*, Fox News, and *National Review*) and three more left-leaning (*Talking Points Memo*, *New York Times*, and *Buzzfeed News*) for our demonstration.

---

[20] One alternative to WMD, and related algorithms, involves taking the average vector of all the words in a document (sometimes called Word Centroid Distance), and comparing these resulting vectors using cosine similarity (Berry & Taylor, 2017; Lix et al., 2020), which is likely well suited for smaller documents, such as social media comments.

[21] Transportation problem uses Earth's Mover's Distance (Rubner et al., 1998) to compare multidimensional distributions. Many have found computationally efficient solutions (Atasu et al., 2017; Tithi & Petrini, 2020; Werner & Laber, 2019; Wu et al., 2018). Our method incorporates one: Linear Complexity Relaxed Word Mover's Distance (LC-RWMD).

[22] "All the News" corpus is a collection of 204,135 news articles from 18 U.S. news organizations, mostly from 2013 to early 2018 (Thompson, 2018).

[23] We collected the press releases by building a scraper using the "rvest" R package (Wickham, 2019).

We find each news organization's similarity to each advocacy organization by the average of the similarity between their respective news articles per year and the press releases pooled by advocacy organization. We show similarities for all articles and also a subset of articles referencing "immigration" or "immigrants" at least once ($N = 4,496$ after also specifying the date range) and plotted their average similarities to the right-wing (red) and left-wing (blue) press releases over time (see Figure 3).

**[FIGURE 3. HERE]**

For left-leaning news organizations, similarity between articles and press releases increases from mid-2016 and peaks in early 2017. When right-leaning news organizations write about "immigration" they are significantly more similar to the press releases from right-wing advocacy organizations than from left-wing advocacy organizations. However, when left-leaning news outlets write about immigration, they are equally similar to both left- and right-wing advocacy organizations, and are also more similar to press releases from either side of the political spectrum, save for early 2016.

This finding aligns with work on echo chambers and political polarization in U.S. society—an important area of contemporary cultural sociology—showing liberals are more likely than conservatives to be comfortable with ambiguity and change, while conservatives shifted more rightward than their liberal counterparts shifted leftward (C. A. Bail et al., 2018; Jost et al., 2007). Specifically, this is evidence that asymmetric polarization (Grossmann & Hopkins, 2016) might be occurring within immigration discourse. If the temporal ordering of these data were such that the press releases were consistently published before the sampled news articles, this question of asymmetric polarization could also be framed as a question of cultural diffusion (C. A. Bail, 2016).

## Document-by-Concept Distances

### Measuring Conceptual Engagement

Concept Mover's Distance (CMD) (Stoltz & Taylor, 2019; Taylor & Stoltz, 2020) quantifies the extent a document "engages with" a theoretically-motivated "focal" concept. To do so, we hold the embedding space constant while measuring the position of documents relative to it, specifically measuring each document's distance from a specific "region" of the embedding space. This can be used to explore how, for example, the linear association between the concept of "death" and actual body counts in Shakespeare's First Folio (Stoltz & Taylor, 2019). Like WMD, CMD relies on a "transportation problem" logic to measure the distance between each document and at least one "pseudo-document."[24]

---

[24] One could measure distance to a larger semantic region by averaging several word vectors forming a "semantic centroid," or using atom topic modeling (Arora et al., 2018; Arseniev-Koehler et al., 2020). In both cases the resulting vector will be in the same dimensional space as the embeddings and can easily be used with the CMD/WMD solver.

In the simplest case, this pseudo-document consists of a single word associated with a single vector in the embeddings. Measuring engagement with more "specific" concepts—e.g., "liberal politics" or "conservative politics" instead of simply "politics"—is accomplished by adding relevant terms to the pseudo-document to create *compound concepts*: for example, by using a pseudo-document consisting of "politics" *and* "liberal" to measure "liberal politics."

Consider again the "All the News" corpus. We use CMD (here with pretrained embeddings), to measure the extent all news articles from January 2012 to March 2018 engaged the concept "immigration." We also measured engagement with the following specified concepts: "immigration + job," "immigration + school," "immigration + crime," and "immigration + family."

Figure 4 shows the smoothed engagement time trends for each concept, averaged by month. According to the plot, engagement with "immigration" and its related compound concepts have followed a similar time trend from 2012 to mid-2018. The "immigration + school" compound concept peaked in late-2014 to mid-2015, around the time the Obama administration's planned extensions to the Deferred Action for Childhood Arrivals (DACA) policy (school requirements were a key element for program eligibility) in late 2014 that were in court battles throughout 2015 and 2016. One such planned extension was the Deferred Action for Parents of Americans and Lawful Permanent Residents (DAPA), aligning with the spike for "immigration + family" during the same time. Whatever the mechanism, this spike in engagement suggests immigration within the contexts of schooling and family seemed to be prevalent modes of media immigration discourse in late 2014 to 2015. Media immigration discourse of all stripes rises in mid-2016 to early-2017—coinciding with Donald Trump winning the Republican primary and the U.S. presidency.

**[FIGURE 4. HERE]**

Figure 5 plots just the monthly average engagement with "immigration" (i.e., the dark purple trend line in Figure 4, with a lower smoothing factor). The figure shows that media engagement rises predictably with relevant sociopolitical events: e.g., introduction DACA policy, the DACA expansion, and Trump's "immigrants-as-rapists" presidential running announcement speech on June 15, 2015. This predictable finding is in line with work on heteronomy in media fields, where fields of journalistic production are known to be highly responsive to shifts and pressures from political (and economic) fields (Benson, 1999; Bourdieu, 2005).

**[FIGURE 5. HERE]**

**Measuring Engagement with Binary Concepts**

Opposing words—i.e. binary concepts—are likely to be near one another within the embeddings space because they are used in similar contexts and they often co-occur (Deese, 1966; Justeson & Katz, 1991; Miller & Charles, 1991, pp. 25–6). As such, oppositions such as "good" and "evil" occupy similar positions in any adequately-trained corpus because they are used in similar ways and mutually oriented toward a shared meaning (Boutyline, 2017; A. Goldberg, 2011). For example, a researcher may be interested in examining engagement with "evil," but a document that engages strictly with "good" would still be highly ranked because the distance of any given word in that document is roughly equidistant to "good" and "evil" within the embeddings space.

To address this, we can use "semantic directions" as described previously (Taylor & Stoltz, 2020). An analyst can measure engagement with one "pole" of a binary by (1) extracting a direction in the meaning space pointing toward a pole of the binary opposition, (2) adding the estimate as a row vector to the embeddings matrix, and (3) adding a pseudo-document to the corpus consisting of only a single reference to that estimated vector. From there, using CMD is the same: it quantifies the cost to move all words in an observed document to the estimated vector—in effect, measuring engagement with one pole of a binary concept as opposed to the other pole (e.g., "evil" *as opposed to* "good").

Consider again the "All the News" corpus. Following the same procedure outlined previously, we constructed two cultural dimensions: an "immigrant" dimension and a "race" dimension, with larger positive values indicating more engagement with the "immigrant" pole as opposed to a "citizen" pole and more engagement with the "black" pole as opposed to the "white" pole, respectively. The immigrant-citizen dimension was constructed using the pairs listed in Table 4, and is understood as measuring a persistent symbolic structure rather than a legal distinction (cf. Beaman, 2016; Jaworsky, 2013).[25]

**[TABLE 4. HERE]**

Smoothed media engagement time trends with the respective poles of these cultural dimensions (averaged by month-year) are shown in Figure 6 top panel; the bottom panel shows how much each month-year's average engagement deviates from the previous month-year's engagement. Engagement with immigration and race cultural dimensions in U.S. news media (specifically, between March 2013 and September 2017) appears to follow similar trends: when news media engage "immigrant" more relative to "citizen" in their discourse, so too do they generally engage "black" more relative "white." This trend is also in line with the above analysis using historical embeddings from 1880 to 2000, again pointing to the remarkable temporal stability of "citizen" and "white" as the unmarked social categories of U.S. public discourses on cultural citizenship and race (Brekhus, 1998) and/or the historical persistence of the "immigrant-as-nonwhite" vs. "citizen-as-white" symbolic boundary (Mora & Paschel, 2020; Sáenz & Douglas, 2015).

---

[25] See footnote 22. While we use "citizen" as the pole opposite "immigrant" to highlight the symbolic structure through which immigrants have been discursively othered in the U.S., citizen is not an antonym for immigrant in any formal way since, obviously, an immigrant can certainly be a legal citizen.

[FIGURE 6. HERE]

## EXTENSIONS

### Synthesized Embedding Spaces

While distinct, variable and fixed embedding space methods could be combined—what one might call *synthesized embedding space* methods. Synthesized approaches would involve simultaneously assessing the changing structure of embedding spaces as a function of some external variable—most likely time—*and also* how documents and authors relate to one another or to a concept vis-à-vis some fixed comparable points across the embedding spaces. For example, if we had a sample of English-language news articles from the past century, we could measure how news organizations spanning several decades engage with a concept or cultural dimension using decade-specific embeddings.

## Beyond "Similarity"

Our demonstration, as well as much current work using embeddings, focuses on semantic *similarity*, which allows us to explore relations of polysemy, synonymy, antonymy, as well as metaphor and metonymy (Erk, 2012). However, there are other relations to be explored: specifically, scale (e.g., good, greater, greatest), classification (i.e., entailment and part-whole relations like hypernymy, hyponymy, meronymy, or holonymy) or object qualities (Fulda et al., 2017; Grand et al., 2018). Most pretrained embeddings already encode these relations, to some extent, and can be extracted with simple post-processing (cf. Fulda et al., 2017; Levy et al., 2015).

For example, one can use "Hearst patterns" (Hearst, 1992)—"such as," "like," "including"—to extract hypernymy/hyponymy (Baroni et al., 2012; Roller & Erk, 2016). Similarly, Fu et al. (2014) propose a post-processing method for learning a linear projection for hierarchical relations using known hypernym/hyponym pairs (see also Roller et al., 2014). Other approaches modify the training phase to better incorporate hierarchical information. For example, Le et al. (2019) use "hyperbolic" space (Nickel & Kiela, 2017) rather than Euclidean space (see also Kruszewski et al., 2015; Weeds et al., 2014). Hyperbolic embeddings with Hearst patterns (Le et al., 2019) might be particularly useful for scholars of culture working with conceptual metaphor theory (Hart-Brinson, 2016; Kharchenkova, 2018; Rotolo, 2020), since models designed to infer transitive hyponym-hypernym relations (e.g., inferring that "conifer *is a* plant" by way of including "conifer *is a* tree" and "tree *is a* plant" in the input Hearst matrix) might also be used to capture the source-target structures behind metaphorical reasoning (e.g., "country *is a* family" (Lakoff, 2010)) that are rarely explicit in written language.

# Concluding Remarks

Formal text analysis in sociology was pioneered by scholars like John Mohr, Wendy Griswold, Karen Cerulo, and Kathleen Carley, and even as it continues to evolve with unprecedented access to large quantities of texts and computational power, the methods of this early era continue to be productive. Although we note limitations, this is not to say count based methods are no longer useful. Rather, embedding methods should be used alongside, as a welcome addition to the cultural analyst's toolbox.

The variety of options available to cultural analysts interested in computational text analysis signals an exciting time. From word counts and dictionary methods to embeddings and other unsupervised learning algorithms, computational text analysis is quickly becoming as institutionalized as ethnography, interviews, and historical-comparative methods. Looking back, it is quite fascinating to see that John Mohr saw this advent coming with such clarity as far back as at least 1998:

> . . . [I]t is probably worth pointing out that we are just now entering what must surely be the golden age of textual analysis. What sets this moment in history apart is the incredible proliferation of on-line and on-disk textual materials. Previously, scholars who were interested in doing some form of content analysis were compelled to spend huge amounts of time readying their texts for analysis. Now one can easily sit at one's desk and more or less instantaneously summon up a fantastic array of cultural texts in electronic form. (Mohr, 1998, p. 366)

We are certainly amid the golden age of text analysis in the social sciences—just as John predicted and helped ensure.

# Appendix

## A. Preprocessing

As our documents (both the news articles and the press releases) were obtained from the Internet, we "preprocessed" our documents by removing non-ASCII characters, URLs, and HTML. We then follow conventional preprocessing steps in the following order: replacing contractions with their full forms using the contractions dictionary in Rinker's "qdapDictionaries" R package (Rinker, 2018), removing punctuation, replacing ordinal numbers with their word form (e.g., "3rd" to "third"), replacing numerals with their word forms (e.g., "3" to "three"), stripping excess whitespace (i.e., spaces between words that are greater than one), and removing capitalization. We also removed terms found in the most commonly used pre-compiled "stop list" (Porter, 2001). This list includes many of the most frequent words in English, e.g., "the," "of," and "and" (for a comparison of several lists, see Nothman et al., 2018). We removed sparse terms at a .999 sparsity factor—meaning terms that were absent in at least 99.9% of the documents. Finally, we removed terms that were not present in both DTMs. For the news articles, this resulted in a DTM of 11,271 unique terms, 79,483,124 total terms, and 197,814 documents. For the press releases, this resulted in a DTM of 11,271 unique terms, 297,977 total terms, and 976 documents. The cell entries in both DTMs indicate the raw count of each retained word in each document (i.e., term frequencies).

## B. Word Embedding Models: SVD, Word2vec, GloVe, and fastText

As stated in the main text, at their most basic, word embedding models involve creating a TCM and reducing its dimensions. This can be accomplished with techniques as common as singular value decomposition (SVD) (Deerwester et al., 1990; Levy et al., 2015, p. 213). Recent advances do roughly the same as SVD while improving the *tuning* of this low-dimensional mapping.[26] Two papers credited with solidifying embedding models as the future of natural language processing were Bengio et al. (2003) and Collobert and Weston (2008); however, it was Mikolov et al. (2013) and the introduction of "word2vec" that popularized these techniques.

Strictly speaking, word2vec does not learn dimensions from the TCM, but rather predicts the word co-occurrence statistics. Word2vec generally refers to two models, one which predicts a target term from context terms, and the other which predicts context terms from a target term. Regarding the former, Levy and Goldberg (2014) found the model was *implicitly* reducing a TCM: the model does not begin with the TCM, but rather extracts co-occurrence statistics as it iterates through the text. This model was soon followed by GloVe (Pennington et al., 2014) and fastText (Joulin et al., 2016), the former of which does *explicitly* reduce the TCM. A primary difference between these models is how they weight very rare or very common words. Additionally, with GloVe models, the individual target-word-by-context-word counts may be weighted by the (inverse of the) distance of the context word to the target word in the context window. Furthermore, fastText (Bojanowski et al., 2017) uses "subword" character $n$-grams and not full words; the vector of a word is the sum of the vectors of the subwords it comprises (see also Schütze, 1993). In particular, this subword approach improves embeddings for rare words and agglutinative languages.

The resulting term vectors are "low-dimensional" in relation to the term co-occurrences of which they are a reduction; however, they still tend to be between 50 and 500 dimensions, with 300 being the most common for pretrained embeddings. Figure 2a in Pennington et al. (2014) shows that accuracy on an analogy task improved up to 300 dimensions for the GloVe model (for a detailed discussion see Spirling & Rodriguez, 2019; Yin & Shen, 2018).

While each approach differs somewhat (for a detailed comparison see Goldberg, 2016), at core they attempt to find a low-dimensional "embedding" space from otherwise high-dimensional term co-occurrences that accurately predicts the context of target words (or vice versa) or similar tasks, such as solving analogies and translations (Lenci, 2018, p. 157; Levy et al., 2015).

Whereas these models summarize the contexts of each unique *token*, recent developments output vectors to represent the potentially distinct "senses" of each token: e.g, *bank* as in river *bank* versus investment *bank*. These "contextualized" word embeddings (Smith, 2019, pp. 6–7), namely ELMo and BERT, are gaining popularity in computational linguistics and information retrieval, but we are only beginning to see how such models might be applied in the social sciences (e.g., Vicinanza et al., 2020). Future work is needed to determine when more elaborate embedding models are preferred over the more straightforward models outlined here (e.g., Dubossarsky et al., 2018). In particular, different "senses" of a token may nevertheless be semantically related (e.g., by analogy or etymology) in ways that should remain connected for the purposes of the analysis.

---

[26] It is outside the scope of this paper to discuss how various researchers measure "accuracy" or "performance" in producing word embeddings, but this is especially a consideration for the cultural analyst wishing to use corpus-trained embeddings (Spirling & Rodriguez, 2019).

## Appendix References

# Tables and Figures

## Tables

**Table 1. Definitions**

| | |
|---|---|
| Word Embeddings | Vector representations of words in an ($n < N$)-dimensional space (see Vector Space Model below), where $N$ is the total number of words in a corpus and vectors are dense (i.e. no zero) and usually consisting of real numbers. |
| Document-Term Matrix (DTM) | A matrix where documents are rows and terms are columns (or vice versa), and a cell entry is a numerical representation of the $j$th word in the $i$th document. |
| Term Co-Occurrence/Context Matrix (TCM) | A symmetric $N$-by-$N$ matrix, where $N$ is the number of terms in a corpus, with cells numerically representing the extent the $j$th term tends to appear in the $i$th term's context window. |
| Relative Frequency | The raw count of the $j$th term in the $i$th document divided by that document's total word count. |
| Word Representation | A discrete number or vector of numbers standing in place of a unique string of characters. |
| Vector Space Model | A representation of words or documents as locations in a "meaning-space" in which semantic relations are understood as geometric relations. |
| Semantic Direction | A vector extracted from an embedding space by subtracting juxtaposing terms (or terms), pointing away from the term (or terms) subtracted. |
| Semantic Region | A vector corresponding to an area of the embedding space built by averaging several related terms (i.e. a semantic centroid) or clustering the embedding space (e.g., atom topic modeling). |

**Table 2. Example of commonly used pretrained embeddings**

|  | Corpora | Tokens | Vectors | Dimensions |
|---|---|---|---|---|
| fastText | Wikipedia 2017, UMBC webbase corpus, statmt.org news | 16 billion | 1 million | 300 |
|  | Common Crawl | 600 billion | 2 million | 300 |
| GloVe | Wikipedia 2014, English Gigaword 5th Edition | 6 billion | 400 thousand | 50 to 300 |
|  | Common Crawl | 840 billion | 2.2 million | 300 |
|  | Twitter | 27 billion | 1.2 million | 25 to 200 |
| word2vec | Google News dataset | 100 billion | 3 million | 300 |

**Table 3. Uses of Variable and Fixed Embedding Spaces**

| | Points of Reference | | |
|---|---|---|---|
| Embedding Space | Terms | Semantic Directions or Semantic Regions | Documents or Authors |
| **Variable**<br><br>A corpus is subset by a covariate (typically by time or author), and embeddings are trained on each subset. | How does the distance between two (or more) terms change from one estimated embedding space to another? | How does the distance between terms and semantic direction or regions change from one estimated embedding space to another? | |
| **Fixed**<br><br>An embedding is trained on a single corpus. Authors or documents are defined as the aggregate of locations of the words associated with them. | How does the distance between documents (or authors) and terms in a single embedding space differ by document covariates? | How does the distance between documents (or authors) and semantic directions or regions in a single embedding space differ by document covariates? | How does the distance between documents (or authors) in a single embedding space differ by document covariates? |

**Table 4. Term Pairs for Immigration-Citizenship Cultural Dimension**

| immigrants | citizens |
|------------|----------|
| immigration | citizenship |
| immigrant | citizen |
| foreign | domestic |
| foreigner | native |
| outsider | insider |
| stranger | local |
| alien | resident |
| foreigner | resident |
| alien | native |
| immigrant | local |
| foreign | familiar |

# Figures

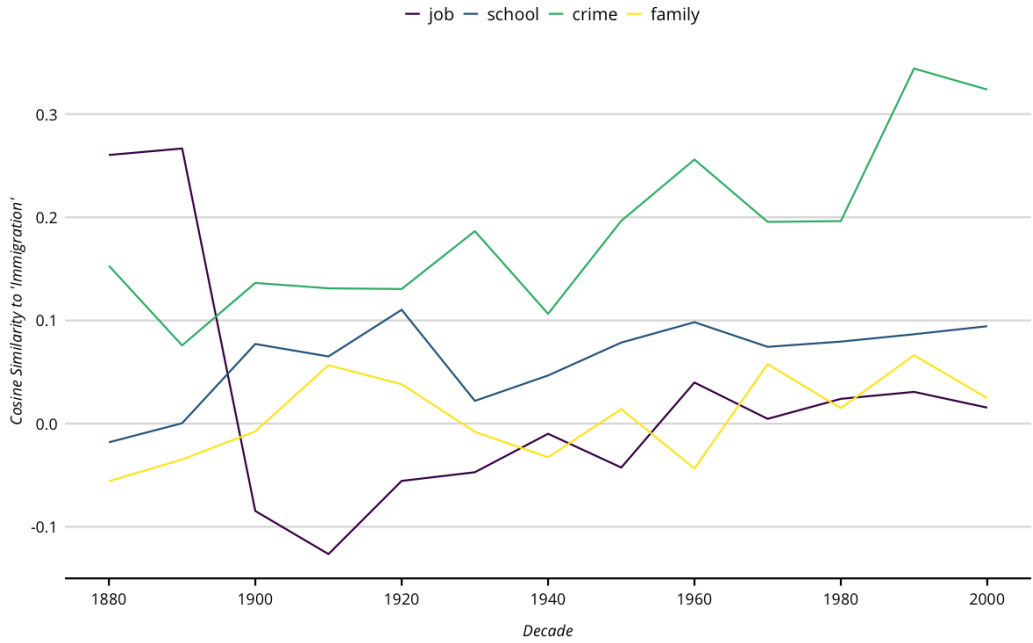

**Figure 1. Cosine Similarity of 'Immigration' and Key Terms by Decade, 1880 to 2000**

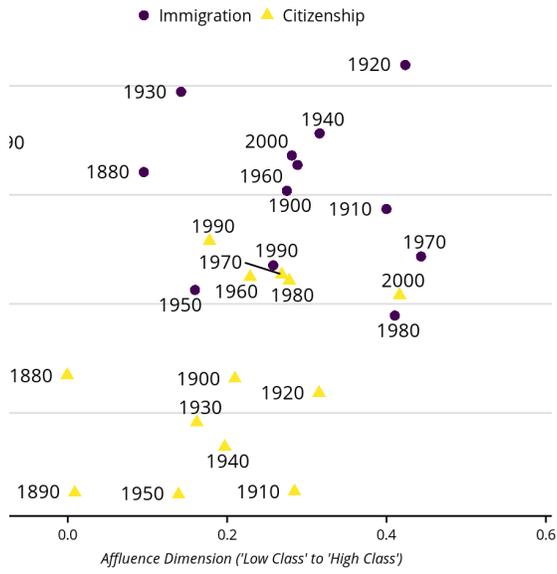

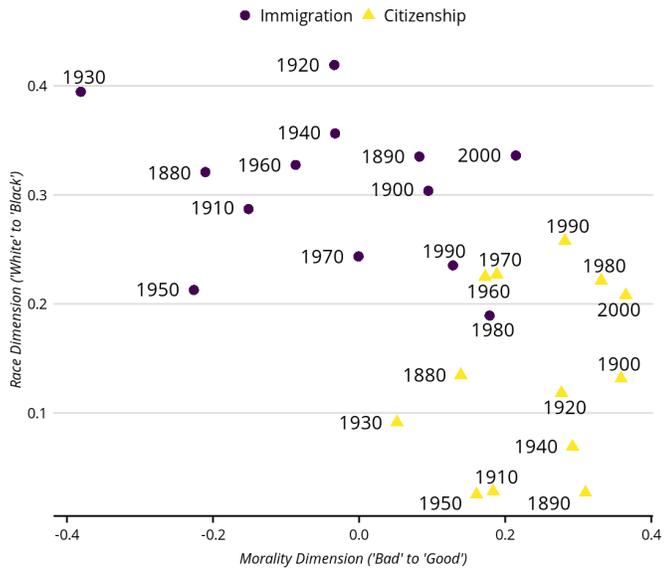

**Figure 2. 'Immigrant' and 'Citizen' on Key Cultural Dimensions, 1880 to 2000**

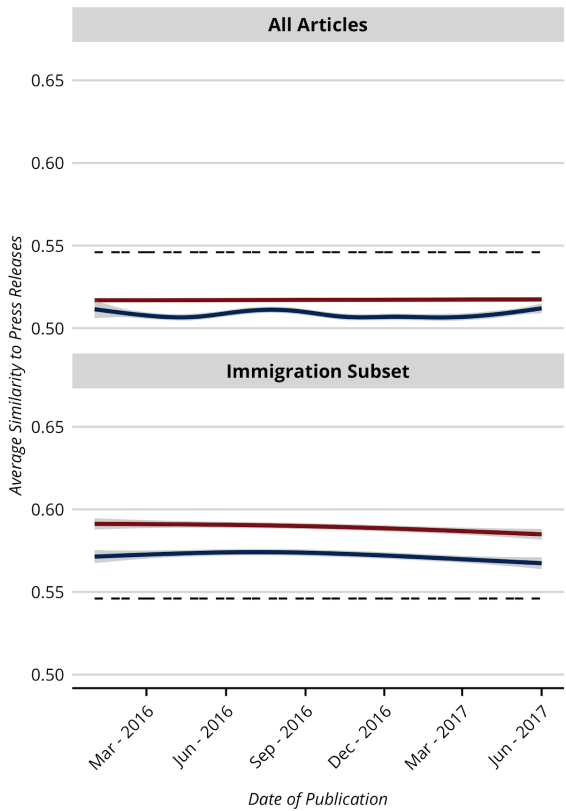
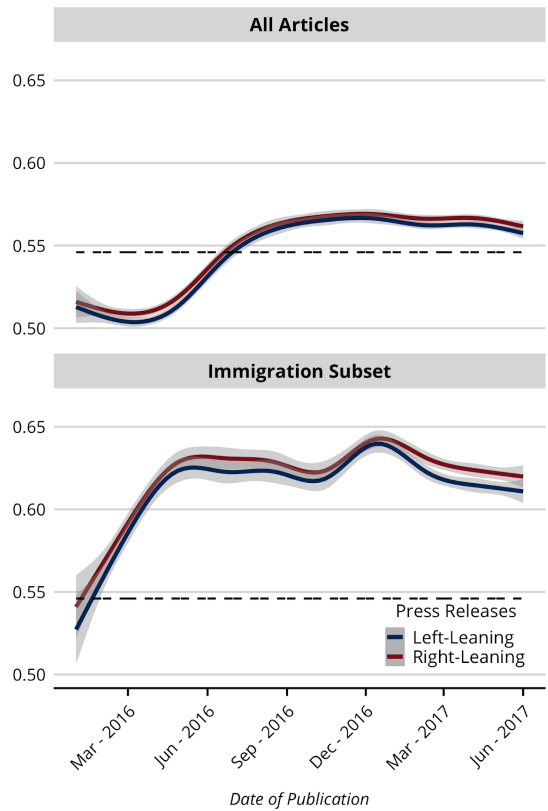

**Figure 3. News Articles' Similarity to Press Releases (with WMD)**

*Note*: Top panels include all articles in the dataset. Bottom panels include only articles referencing "immigration" or "immigrants" at least once. The dashed line represents the mean similarity to the press releases for all news articles, irrespective of political lean.

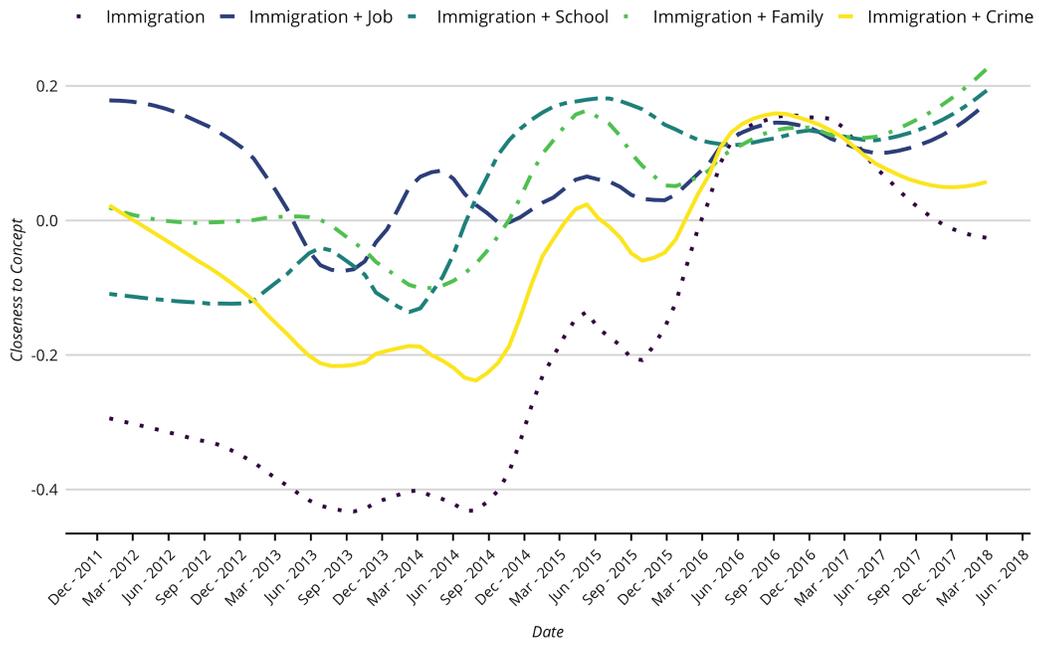

**Figure 4. News Articles' Conceptual Engagement Over Time (with CMD)**

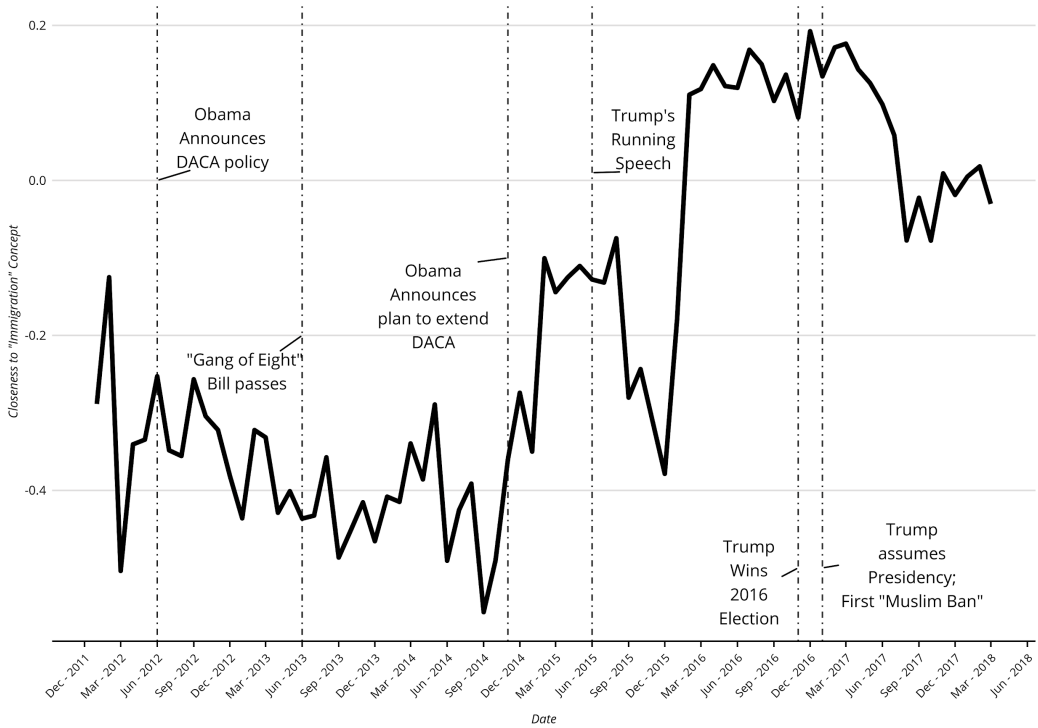

**Figure 5. News Articles' Conceptual Engagement and Key Events (with CMD)**

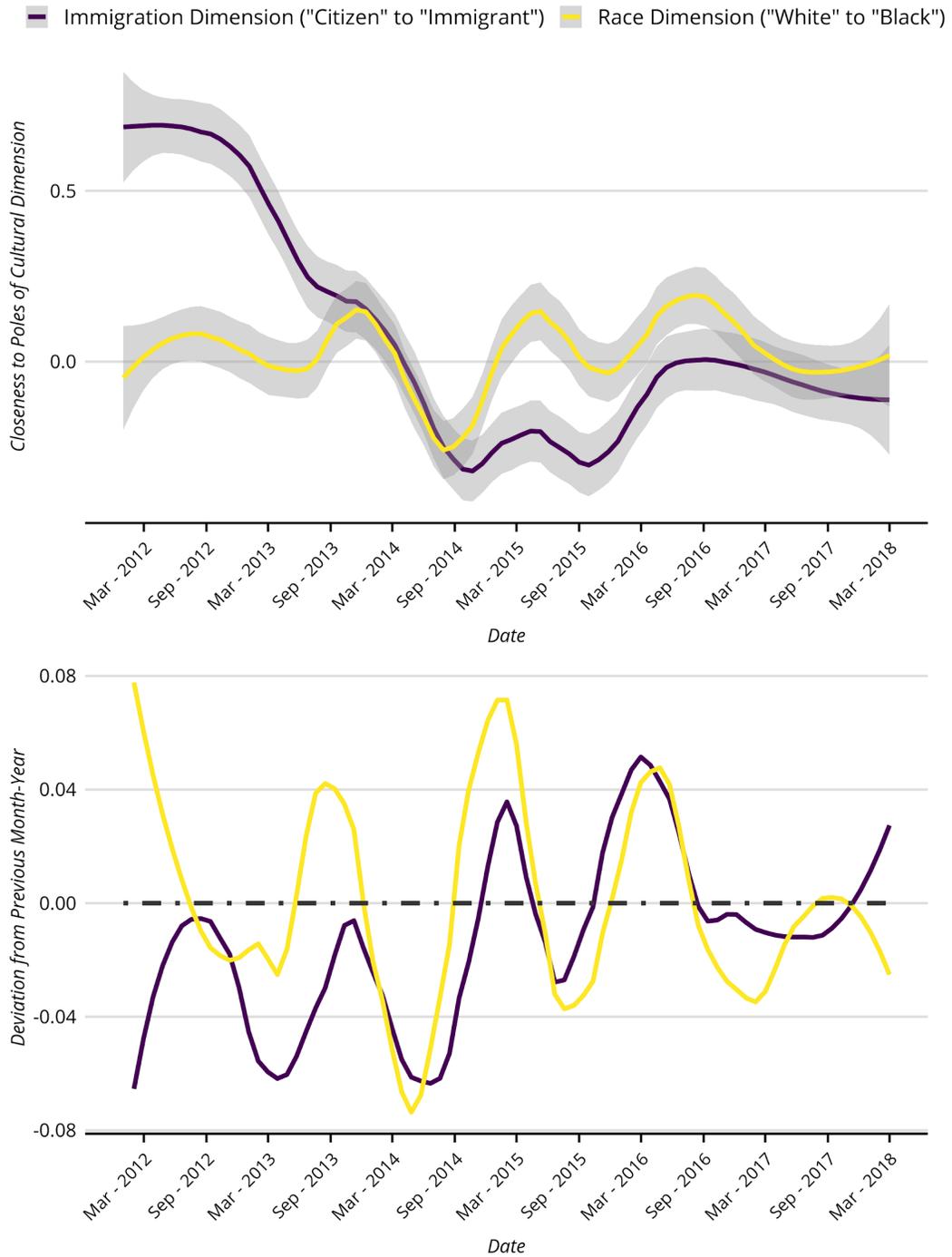

**Figure 6. News Articles' Engagement with Key Cultural Dimensions (with CMD)**